\documentclass[%
reprint,
groupedaddress,
openany
bibnotes,
 amsmath,amssymb,
 aps,
prc,
]{revtex4-1}

\usepackage{graphicx}% Include figure files
\usepackage{epstopdf}
\usepackage{dcolumn}% Align table columns on decimal point
\usepackage{bm}% bold math
\begin{document}

\preprint{APS/123-QED}

\title{Aspect of Clusters Correlation at Light Nuclei Excited State}
\email{tfwang@buaa.edu.cn (T. F. Wang)}
 \author{Ziming Li$^{1}$, Jie Zhu$^{1}$, Taofeng Wang$^{1*}$,  Minliang Liu$^{2}$, Jiansong Wang$^{2}$,\\ Yanyun Yang$^{2}$, Chengjian Lin$^{3}$, Zhiyu Sun$^{2}$, Qinghua He$^{4}$, M. Assié$^{5}$,Y. Ayyad$^{6}$,\\ D. Beaumel$^{5}$, Zhen Bai$^{2}$, Fangfang Duan$^{2}$, Zhihao Gao$^{2}$, Song Guo$^{2}$, Yue Hu$^{1}$,\\ Wei Jiang$^{7}$,  F. Kobayashi$^{8}$, Chengui Lu$^{2}$, Junbing Ma$^{2}$, Peng Ma$^{2}$,\\ P. Napolitani$^{9}$, G. Verde$^{10,11}$, Jianguo Wang$^{2}$, Xianglun Wei$^{2}$, Guoqing Xiao$^{2}$,\\ Hushan Xu$^{2}$, Biao Yang$^{7}$, Runhe Yang$^{2}$, Yongjin Yao$^{1}$, Chaoyue Yu$^{2}$,\\ Junwei Zhang$^{2}$, Xing Zhang$^{2}$, Yuhu Zhang$^{2}$,  Xiaohong Zhou$^{2}$}
\affiliation{%
 $^{1}$School of Physics, Beihang University, Beijing 100191, China\\ 
 $^{2}$Institute of Modern Physics, Chinese Academy of Sciences, Lanzhou 730000, China\\
 $^{3}$China Institute of Atomic Energy, P.O. Box 275 (10), Beijing 102413, China\\
 $^{4}$Department of Nuclear Science $\&$ Engineering, College of Material Science and Technology,\\Nanjing University of Aeronautics and Astronautics, Nanjing 210016, China\\ 
 $^{5}$IJCLab, Université Paris-Saclay, CNRS/IN2P3, 91405 Orsay, France\\
 $^{6}$Facility for Rare Isotope Beams, Michigan State University, East Lansing, Michigan 48824, USA\\
 $^{7}$State Key Laboratory of Nuclear Physics and Technology, School of Physics, Peking University, Beijing 100871, China\\
 $^{8}$Graduate School of Engineering Science, Osaka University, 1-3 Machikaneyama, Toyonaka, Osaka 560-8531, Japan\\
 $^{9}$IPN, CNRS/IN2P3, Université Paris-Sud 11, Université Paris-Saclay, 91406 Orsay Cedex, France\\
 $^{10}$INFN Sezione di Catania, via Santa Sofia 64, I-95123 Catania, Italy\\
 $^{11}$Laboratoire des 2 Infinis - Toulouse (L2IT-IN2P3), Université de Toulouse, CNRS, UPS, F-31062 Toulouse Cedex 9, France
  }
%\date{\today}% It is always \today, today,
             %  but any date may be explicitly specified

\begin{abstract}
The correlation of $\alpha\alpha$ was probed via measuring the transverse momentum $p_{T}$ and width $\delta p_{T}$ of one $\alpha$, for the first time, which represents the spatial and dynamical essentialities of the initial coupling state in $^{8}$Be nucleus. The weighted interaction vertex of 3$\alpha$ reflected by the magnitudes of their relative momentums and relative emission angles proves the isosceles triangle configuration for 3$\alpha$ at the high excited energy analogous Hoyle states.

%\begin{description}
%\item[PACS numbers]
%25.30.Fj, 13.60.Hb, 24.85.+p

%\end{description}
\end{abstract}

\pacs{Valid PACS appear here}% PACS, the Physics and Astronomy
                             % Classification Scheme.
%\keywords{Suggested keywords}%Use showkeys class option if keyword
                              %display desired
\maketitle

%\tableofcontents

\section{Introduction}
Cluster structures are common at all scales from the micro scope of two-nucleon short range correlated pair [1-5] as well as the compact binding four-nucleon $\alpha$ cluster [6] dominated by the strong interaction inside nuclei, to the macro scope of the common clouds, raindrops and snowflakes with the cluster structure dominated by the electromagnetic interaction, and to the cosmological scope of galaxy clusters dominated by the gravity. All levels of matter systems can be regarded as the cluster structures based on the broad interactions.

In the studies of various astrophysical scenarios, the reaction rate of triple-$\alpha$ reaction is crucial for determining the carbon and oxygen abundances at the end of helium burning, with important effects for both nuclearsynthesis and late-stage stellar evolution [7]. The dominant charged particle reaction process, which is primarily induced by $\alpha$ particle fusion in the formation of the elements with atomic mass number $A$ less than that of iron, occurs dependening on the probability overlap between the gravitational potential energy conversion for thermal kinetic energy and the Coulomb barrier penetration at the enough high temperature of 1-2$\times$10$^{8}$ K. Under this condition there will be a small equilibrium concentration of $^{8}$Be ($\alpha$ + $\alpha$ $\leftrightarrow$ $^{8}$Be) with the order of the Boltzmann factor of 4$\times$10$^{-3}$ of the $\alpha$ particle  thermal distribution.

The 2$^{+}$ state of nucleus $^{8}$Be$^{\ast}$ is unstable to decay as two $\alpha$, the short lifetime of which hinders the subsequent synthesis of heavier elements such as $^{12}$C and $^{16}$O.  It was believed [7, 8] that carbon in the universe is in the core of the star through triple-$\alpha$ process synthesis, namely two $\alpha$ firstly forming an ground state of $^{8}$Be with a lifetime of only 10$^{-16}$ s, and before this $^{8}$Be  decay it fuse another $\alpha$ to form the excited states of $^{12}$C ($^{8}$Be + $\alpha$ $\rightarrow$ $^{12}$C$^{\ast}$) and then release $\gamma$-ray as deexcitation. There exists a resonance state of 7.65 MeV above the 3$\alpha$ breakdown threshold with the spin and parity of 0$^{+}$ in $^{12}$C named Hoyle state [9], which improves the reaction rate of $^{8}$Be fusing one $\alpha$.   

It is essential in details to search if there exists the analogous Hoyle states over 7.65 MeV with 3$\alpha$ configuration to enhance the probability of $^{8}$Be fusion one $\alpha$ to form $^{12}$C. This fusion process, known as the triple-alpha process, is crucial for the production of carbon in stars and provides a bridge to the synthesis of heavier elements. If the energy of the analogue Hoyle state were even slightly different, the carbon production in stars would be significantly hindered. 

In order to make clarification for these high excited analogous Hoyle states which may be reconstructed by three detected $\alpha$ from the decay of $^{12}$C$^{\ast}$ formed by $\alpha$ transfer or fusion process, it is necessary to perform the high resolution measurements  with a broad acceptance in the forward angle region using a zero-degree charged particle detection spectrometer.

The two identical particles correlation depending on relative momentums and open emission angles are essential to understand the nature of the dynamical, lifetime and spatial size of the initial state for the particle system consisting of identical constituents. The two identical nucleons of $nn$ pair or $pp$ pair were probed to correlatively existing in nuclei, while no bound state of them exist in the free space. It is crucial to study the correlation of $\alpha\alpha$ which reflects the spatial-temporal properties of $^{8}$Be produced from two-$\alpha$ process. These time and space aspects of $^{8}$Be essentially impact the subsequent triple-$\alpha$ process for forming $^{12}$C, and play a key role in the helium burning process and the carbon-nitrogen-oxygen (C-N-O) cycle in the primordial epoch of the elements production [9-13].

For the spatial structure of 3$\alpha$ at the Hoyle state, once there are different opinions about the linear form, equilateral triangle and $\alpha$ condensation with a dilute gas state and so on [12]. H. Morinaga analyzed the excited state of the $\alpha$ conjugated nucleus, and believed that three $\alpha$ of Hoyle resonance state of $^{12}$C are in a straight line structure [10]. So far, no convinced conclusion has been exactly reached for this intrinsic problem. It is extremely attractive to make clear for the spatial structure of 3$\alpha$ at analogous Hoyle states of $^{12}$C.

In this article, we present that the $\alpha\alpha$ correlation by the the transverse momentums $p_{T}$ and widths $\delta p_{T}$ of $\alpha$ in the correlated $\alpha+\alpha$ cluster configurations at the excited $^{8}$Be nucleus, for the first time. Remarkably, the isosceles triangle configuration of 3$\alpha$ in the analogous Hoyle states is experimently ascertained by measuring the correlation of relative momentums and open emission angles of each $\alpha_{i}\alpha_{j}$ pair. 

\begin{figure}[htb]
\includegraphics[width=8.0cm]{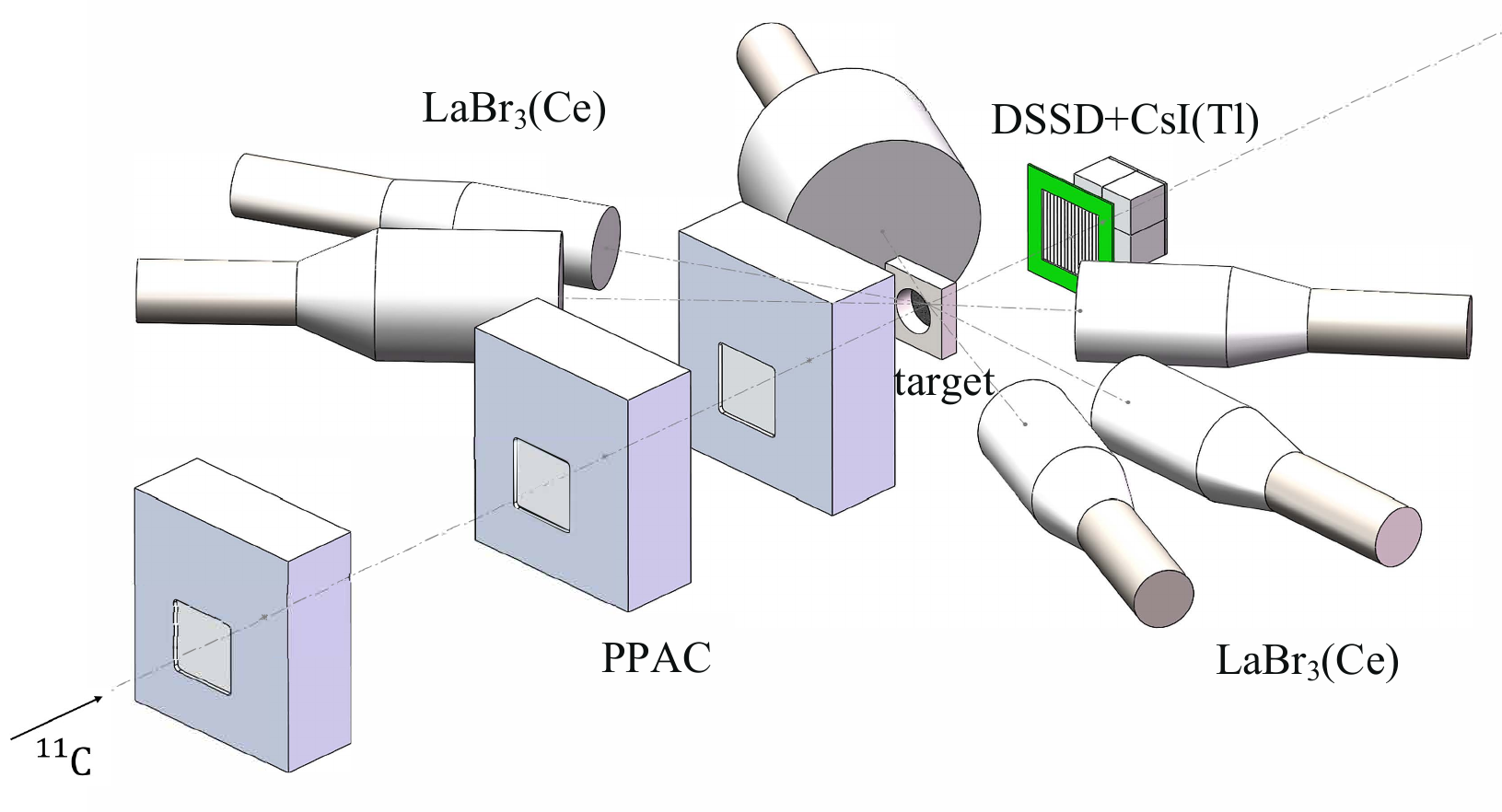}% Here is how to import EPS art
\caption{\label{fig:pdfart} The experimental equipment consisting of three PPACs for determing the reaction position of beam particle on the target, the reaction products were detected by a DSSD silicon detector combined with a 2$\times$2 CsI(Tl) scintillators array, the decay $\gamma$s were detected by five LaBr$_{3}$ (Ce) and one NaI scintillator detectors.}
\end{figure}

\begin{figure}
\includegraphics[width=5.5cm]{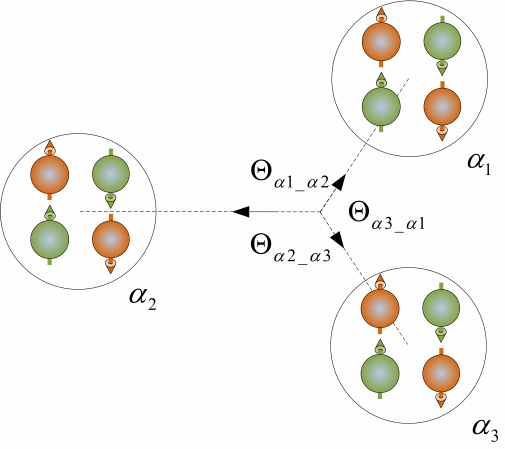}% Here is how to import EPS art
\caption{\label{fig:pdfart}Three $\alpha$ isosceles triangle configuration formed by the excited states of $^{12}$C ($^{8}$Be + $\alpha$ $\rightarrow$ $^{12}$C$^{\ast}$) reaction process. The green and orange balls denote protons and neutrons as constituents of $\alpha$ particle, respectively.}
\end{figure}

\section{Experimental procedure}
The present experimental measurement was performed at the Radioactive Ion Beam Line at the Heavy Ion Research Facility in Lanzhou (HIRFL-RIBLL) [14, 15]. A 60 MeV/nucleon $^{12}$C beam was used to bombard 3.5 mm $^{9}$Be target to produce about 25 MeV/nucleon $^{11}$C secondary beam with a purity of about 99$\%$ and an intensity of about 10$^{4}$ particles per second. The beam particles were identified in terms of B$\rho$-TOF-$\Delta$E method [14]. The $^{11}$C secondary beam were bombarded on a 50 mg/cm$^{2}$ carbon target to produce the excited $^{8}$Be fragments due to the breakup reaction and the excited $^{12}$C nucleus from the one $\alpha$ exchange reaction.

Three parallel plate avalanche chambers (PPACs) with 50$\times$50 mm$^{2}$ active area and position resolutions of about 1 mm (FWHM) in both the $x$ and $y$ directions were placed in front of the target to track the incident $^{11}$C beam and to subsequently get the reaction vertex in the target. $\alpha$ particle and other products from the excited fragments decay were detected by a zero-degree telescope system which consists of a double-sided silicon strip detector (DSSD, of 148 $\mu$m in thickness and 50$\times$50 mm$^{2}$ in cross-sectional area) with 32 strips on both front and back sides, and a 2$\times$2 photodiode (PD) readout CsI (Tl) scintillator (25$\times$25$\times$30 mm$^{3}$ size for each unit) array. Each CsI (Tl) scintillator is covered by two layers of high reflection Tyvek papers and a 10 $\mu$m aluminum coated Mylar film as window. The PD is coupled to CsI (Tl) scintillator with the photoconductive silicone grease. The angular coverage of the zero-degree telescope is about 0-9$^{\circ}$. Five LaBr$_{3}$ (Ce) and one NaI scintillator detectors were placed around the target to measure the decayed $\gamma$s from the excited fragments. DSSD was utilized to record the $\Delta$E energy and the position of the detecting particles, therefore, the emission angle may be obtained by combining with the reaction vertex in the target. CsI (Tl) detection system provides the residual E energy of the charged particles from fragments decay. Particle identifications (PID) were performed using $\Delta$E-E contour. The energy resolution of this zero-degree $\Delta$E-E telescope detection system was estimated to be $\sim$0.8 MeV from the numerical simulation.

Energy calibration of DSSD was carried out with the $\alpha$ radioactive sources as well as $^{6}$Li, $^{7}$Li, $^{7}$Be and $^{10}$B ions produced from $^{12}$C  bombarding $^9$Be target  and selected at three different magnetic rigidity settings, while CsI (Tl) detectors were calibrated with only those ions. PPACs were calibrated by the time information own to the delay readout method. LaBr$_{3}$ (Ce) scintillator detectors were calibrated with a $^{152}$Eu $\gamma$ source and its own $\alpha$ background.

\section{Data analysis procedure}

\subsection{Single level Breit-Wigner formalism in R-matrix framework}
The resonance properties of a nucleus can be described by various parameters, such as the energy, total and partial widths, and spin-parity values. These properties dictate the behavior of the nucleus during scattering experiments or nuclear reactions involving that particular resonance. The R-matrix formalism is a mathematical framework utilized to model the interactions between particles in the nucleus and external particles or projectiles [16].

The R-matrix approach involves dividing the nuclear system into two distinct regions: the internal region, which encompasses the resonant states, and the external region, consisting of the non-resonant background. The R-matrix fit aims to determine the parameters characterizing the resonant states by fitting theoretical predictions to the experimental data. This fitting procedure involves adjusting the parameters within the R-matrix formalism to best reproduce the observed scattering or reaction cross-sections.

In the R-matrix formalism, the cross-section for a nuclear reaction is related to the R-matrix element, which characterizes the scattering process. The R-matrix element can be written as a sum of terms, each corresponding to a resonance state in the system. Each resonance state is described by its energy, width, and couplings to different reaction channels.

The single level Breit-Wigner formula, on the other hand, provides a convenient mathematical representation for the cross-section associated with a single resonance state. It assumes that the resonance can be described by a Lorentzian shape in the energy domain, where the maximum cross-section corresponds to the resonance energy.

\begin{equation}
\begin{split}
\sigma(a,b)=\frac{2I_{C}+1}{(2I_{a}+1)(2I_{A}+1)}\pi\lambdabar^{2}\frac{\Gamma_{a}\Gamma_{b}}{(E'-E_{0})^{2}+(\Gamma/2)^{2}},
\end{split}
\end{equation}

where $I_{a}$, $I_{A}$ and $I_{C}$ are the spins of incident particle, terget nucleus and compound nucleus. $\lambdabar$=$\lambda/2\pi$ is De Broglie wavelength, $\Gamma_{a,b}$ and $\Gamma$ are the partial and total width, $E'$ and $E_{0}$ are the energy and resonance energy.

A common approach to correct the effects due to the experiemntal resolution including beam width, detector resolution and target thickness spreading is given by the convolution function [17]

\begin{equation}
\begin{split}
F(E_{0})=\int_{E_{0}-\Delta}^{E_{0}}\int_{E=-\infty}^{+\infty}\frac{\sigma(E')}{\epsilon(E')}g(E-E_{0})dE'dE,
\end{split}
\end{equation}

where $\sigma(E')$ is the cross section devoid of experimental effects. The function $g(E-E_{0})$ is the spreading function that represents the energy distribution of the experimental effects [17].

\begin{equation}
\begin{split}
g(E-E_{0})=\frac{1}{\sqrt{2\pi}}\sigma_{e}exp[ -\frac{(E-E_{0})}{2\sigma_{e}^{2}}],
\end{split}
\end{equation}

where $\sigma_{e}$ defines the energy width (sigma) of the experimental effects.

The connection between R-matrix fitting and the single level Breit-Wigner formula lies in the fact that the R-matrix boundary condition can be used to derive the Breit-Wigner expression for the scattering cross-section. By appropriately choosing the R-matrix parameters, one can reproduce the behavior of the cross-section as described by the Breit-Wigner formula.

To perform a single level Breit-Wigner fit within the R-matrix formalism, one assumes a single resonance hypothesis and selects a resonance energy, width, and couplings. These parameters are then adjusted iteratively to minimize the difference between the calculated cross-section given by the R-matrix and the experimental data.

\subsection{$\alpha$-$\alpha$ correlation in $^{8}$Be}

\begin{figure}[htb]
\includegraphics[width=9.2cm]{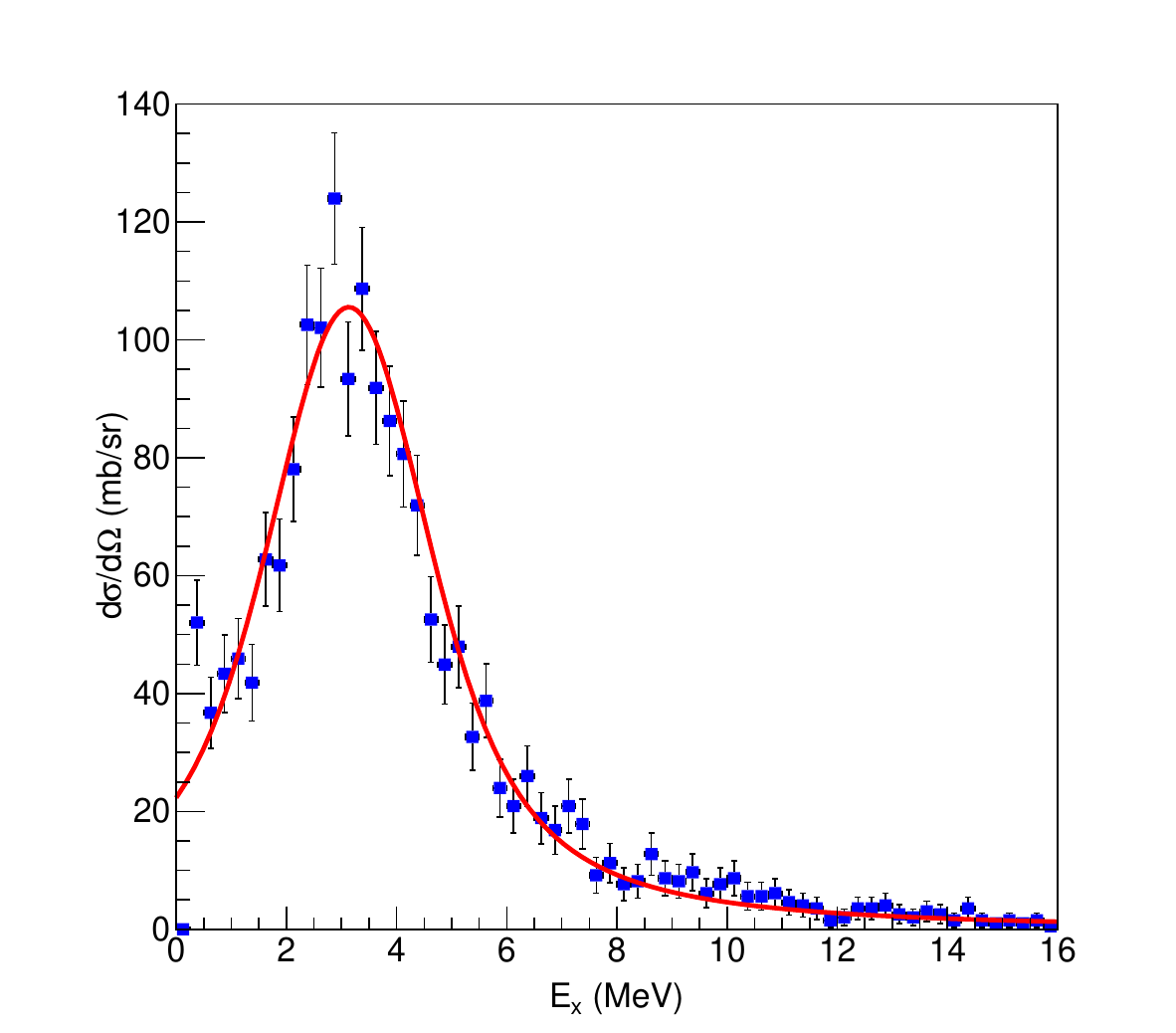}% Here is how to import EPS art
\caption{\label{fig:pdfart} The 2$^{+}$ state of the excitation energy spectrum of $^{8}$Be$^\ast$ reconstructed from the invariant mass of two coincident $\alpha$s, which was fitted by a convolution of Gaussian and Breit-Wigner function. }
\end{figure}

The excited states of $^{8}$Be$^{\ast}$ was reconstructed by the invariant mass $M^{\ast}$ from final state particles of $\alpha$+$\alpha$ selected with the multiplicity-2 events recorded by the zero-degree telescope system. 

\begin{equation}
\begin{split}
M_{^{8}Be^{\ast}}c^{2}=[(\sum m_{i}c^{2}+\sum T_{i})^{2}-(|\sum \vec{p_{i}}|c)^{2}]^{1/2},
\end{split}
\end{equation}
where $m_{i}$, $T_{i}$ and $\vec{p_{i}}$ are the mass, kinetic energy and momentum of two $\alpha$. The excitation energy $E_{x}$ is then given by:

\begin{equation}
\begin{split}
E_{x}=M_{^{8}Be^{\ast}}c^{2}-M_{0}c^{2},
\end{split}
\end{equation}
where $M_{0}$ is the rest mass of the groundstate $^{8}$Be.

The 2$^{+}$ (3.09 MeV) excited state of $^{8}$Be reconstructed from the invariant mass of measured two coincident $\alpha$ particles was fitted by a convolution function of Gaussian and Breit-Wigner expression, as shown in Equ. (1)-(3). None other resonances are observed in the spectrum below 16 MeV, which indicates 2$^{+}$ state existing as the only final state of 2$\alpha$ process to form $^{8}$Be$^{\ast}$. It is convenient to perform such fitting for 2$^{+}$ state of $^{8}$Be by using this convolution function of Gaussian combined to the single level Breit-Wigner expression. The resonance energy of 2$^{+}$ excited state is fitted of 3.09$\pm$0.05 MeV, and the width $\Gamma$ of the 2$^{+}$ excited state from the fitting is 3.20$\pm$0.13 MeV with excluding the experimental setup resolution, the width quantity is larger than the compilated data for the astrophysical scenarios models [18].

The open emission angle distribution of two $\alpha$ forming $^{8}$Be$^{\ast}$ in its center of mass systerm is shown in Fig. 4, where the back-to-back emissions reflect well the characteristics of the two-body decay from the initial state of $^{8}$Be$^{\ast}$. The M-C simulation with the setting of two $\alpha$s emission by the relative angle of 180$^{\circ}$ in the center of mass system after smearing, reproduce the measurement results well in Fig. 4.

\begin{figure}[htb]
\includegraphics[width=9.2cm]{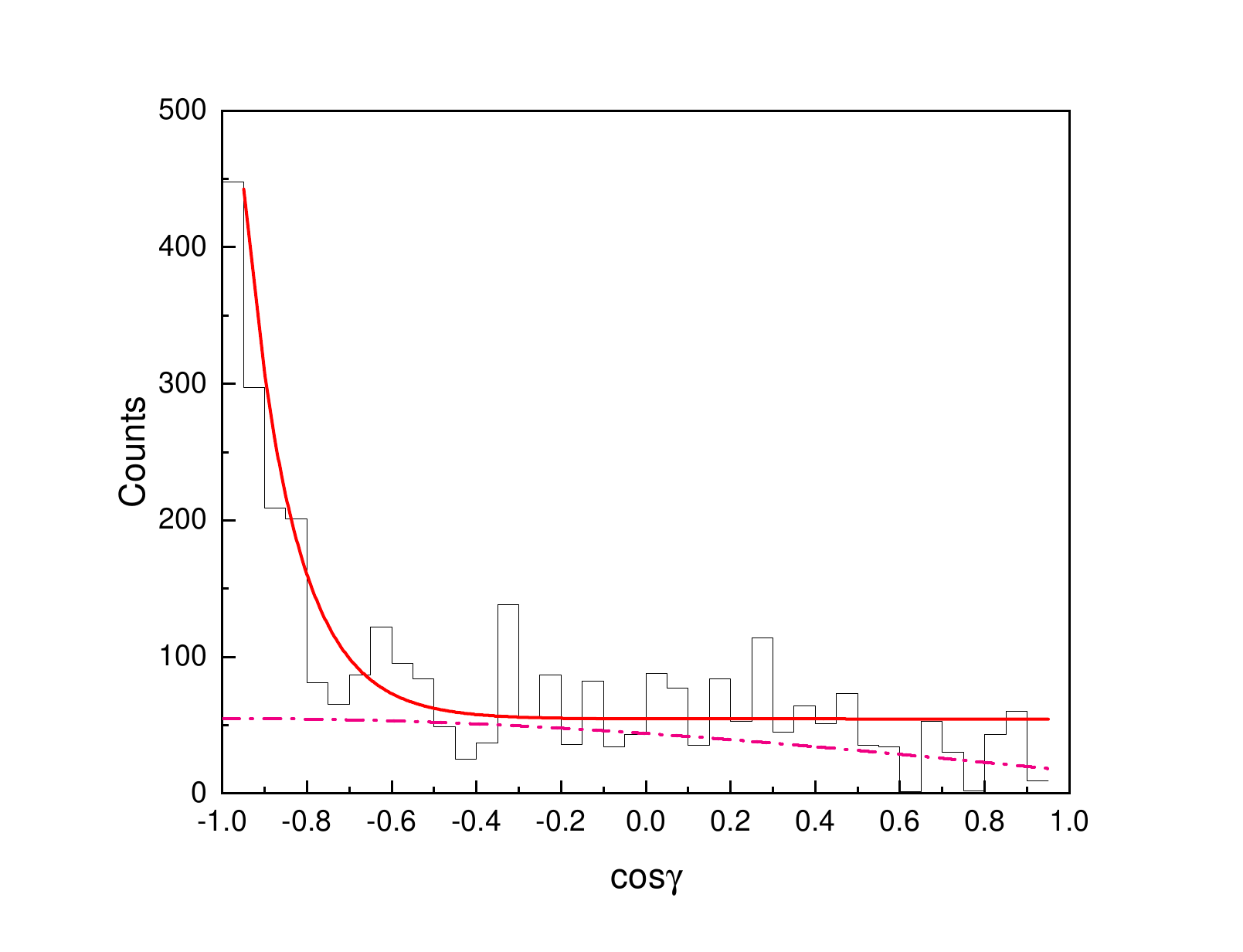}% Here is how to import EPS art
\caption{\label{fig:pdfart} The cos$\gamma$ distribution of two $\alpha$ open emission angle in the center of mass systerm of $^{8}$Be$^{\ast}$. The solid curve exhibits the M-C simulation and the dash dot line shows the unitary level of the random background.}
\end{figure}

The correlation function of $\alpha\alpha$ and $\alpha$ with the relative momentum $p_{rel}$ in the lab systerm, $C_{\alpha\alpha}(p_{rel})=\frac{N_{c}(p_{rel})}{N_{nc}(p_{rel})}$, as shown in Fig. 5,  where multiplicity-2 events were selected for both correlated $\alpha\alpha$ yield of $N_{c}$ and uncorrelated two $\alpha$s yield of $N_{nc}$. The correlation function was decomposed by two Gaussian functions with $p_{rel_{1}}$=48.11$\pm$2.37 MeV/c, $p_{rel_{2}}$=239.76$\pm$6.32 MeV/c and $\sigma _{1}$=17.50$\pm$2.70 MeV/c, $\sigma _{2}$=150.05$\pm$7.86 MeV/c for the mean and sigma values, which corresponse to the ground state and 2$^{+}$ state of $^{8}$Be. The seperated two peaks result from the nautre of single nucleon formed mean field for the ground state and the $\alpha$$\alpha$ cluster correlation of 2$^{+}$ state of $^{8}$Be, which also indicate the small affects from the event-mixing to the shape of the invariant mass spectrum of 2$^{+}$ state in Fig. 4.

\begin{figure}
\includegraphics[width=9.2cm]{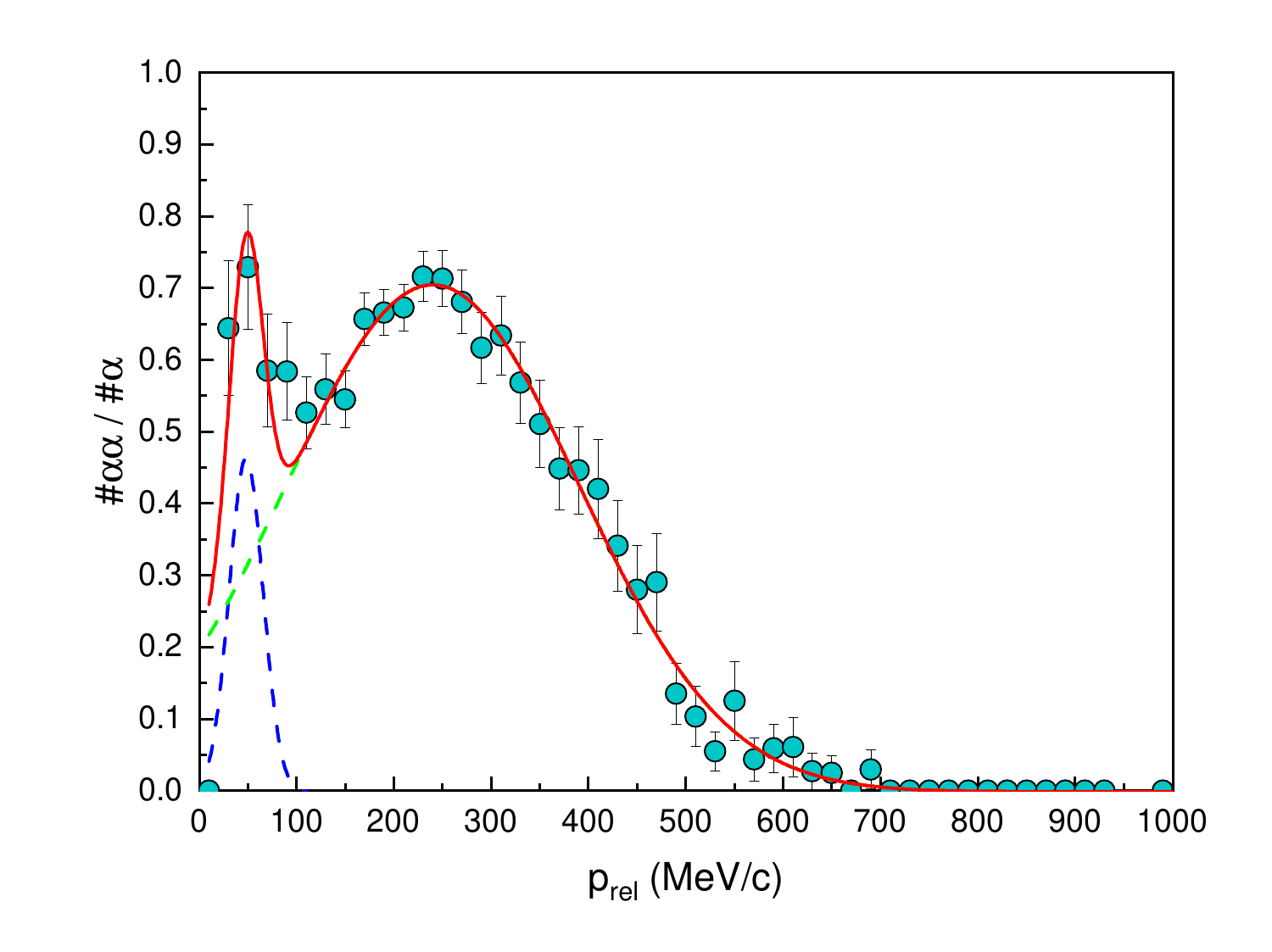}% Here is how to import EPS art
\caption{\label{fig:pdfart} The correlation function $C_{\alpha\alpha}(p_{rel})$ of $\alpha\alpha$ and $\alpha$ with the relative momentum $p_{rel}$ in the lab systerm,  which was decomposed by Gaussian functions.}
\end{figure}

The transverse momentum $p_{T}$ of each $\alpha$ particle in the center of mass system reflects the dynamical motion of $\alpha$ clusters in the initial state of $^{8}$Be$^{\ast}$ before decays. The mean value and width (1$\sigma$) of the transverse momentum distribution of $\alpha$ in $^{8}$Be are 75.03$\pm$1.02 MeV/c and $28.75\pm$0.35 MeV/c, as shown in Fig. 6. The much smaller width of $\alpha$ transverse momentum distribution in $^{8}$Be$^{\ast}$ represents the loosely bound two $\alpha$ cluster structure in 2$^{+}$ state.

The momentum distribution of $\alpha$ in $^{8}$Be at the excited state of $J^{\pi}=2^{+}$ and the ground state of 0$^{+}$ were calculated by the many-body variational Monte Carlo (VMC) [19] with the phenomenological two-nucleon and three-nucleon potential. We calculated the averaged relative $\alpha$ momentum from the VMC simulations, which are conveniently close to the present experimental measurement for the total relative momentum $\vec p_{rel}=\vec p_{\alpha}-\vec p_{\alpha}$ of two $\alpha$ at the lab system shown in Table 1. The averaged $\alpha$ momentum at excited state $2^{+}$ is quite larger than that at the ground state $0^{+}$. This can be understood from the phenomenological configuration that the ground state of $0^{+}$ mainly consists of seperated neutrons and protons forming the nucleon mean-field, while $2^{+}$ excited state takes on two $\alpha$ cluster structure with a highly relative momentum.

\begin{figure}
\includegraphics[width=9.2cm]{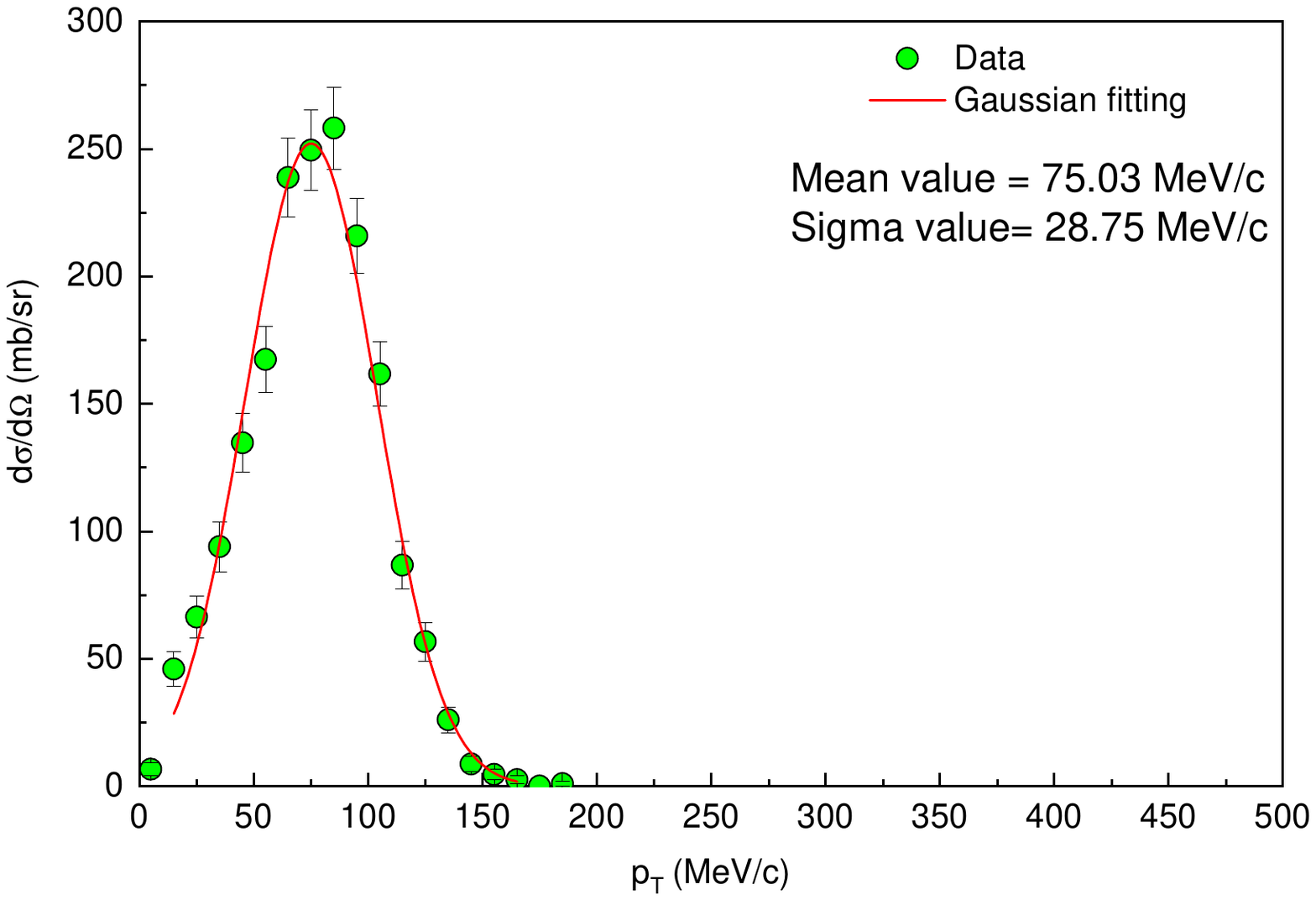}% Here is how to import EPS art
\caption{\label{fig:pdfart} Transverse momentum distribution of $\alpha$ in the center of mass systerm of $\alpha$+$\alpha$ from $^{8}$Be$^{\ast}$ decay, which was fitted by Gaussian functions.}
\end{figure}

\begin{table}
\caption{\label{tab:table1}  Relative $\alpha$ momentum $p_{rel}$ in the lab systerm of the present measurement together with VMC calculation for the excited state $2^{+}$ and ground state $0^{+}$ in $^{8}$Be$^{\ast}$.}
\footnotesize
\begin{ruledtabular}
\begin{tabular}{ccccc}
 &    Present data&  & VMC calculation&   \\
\hline
$J^{\pi}$  & $2^{+}$  &     & $2^{+}$,    $0^{+}$\\
$p_{rel}$(MeV/c)   & 215.59$\pm$1.22 &   & 240.68, 31.69\\
$\delta p_{rel}$(MeV/c)   & 51.20$\pm$1.35 &   & -, -\\
\end{tabular}
\end{ruledtabular}
\end{table}

\begin{figure}[htb]
\includegraphics[width=9.2cm]{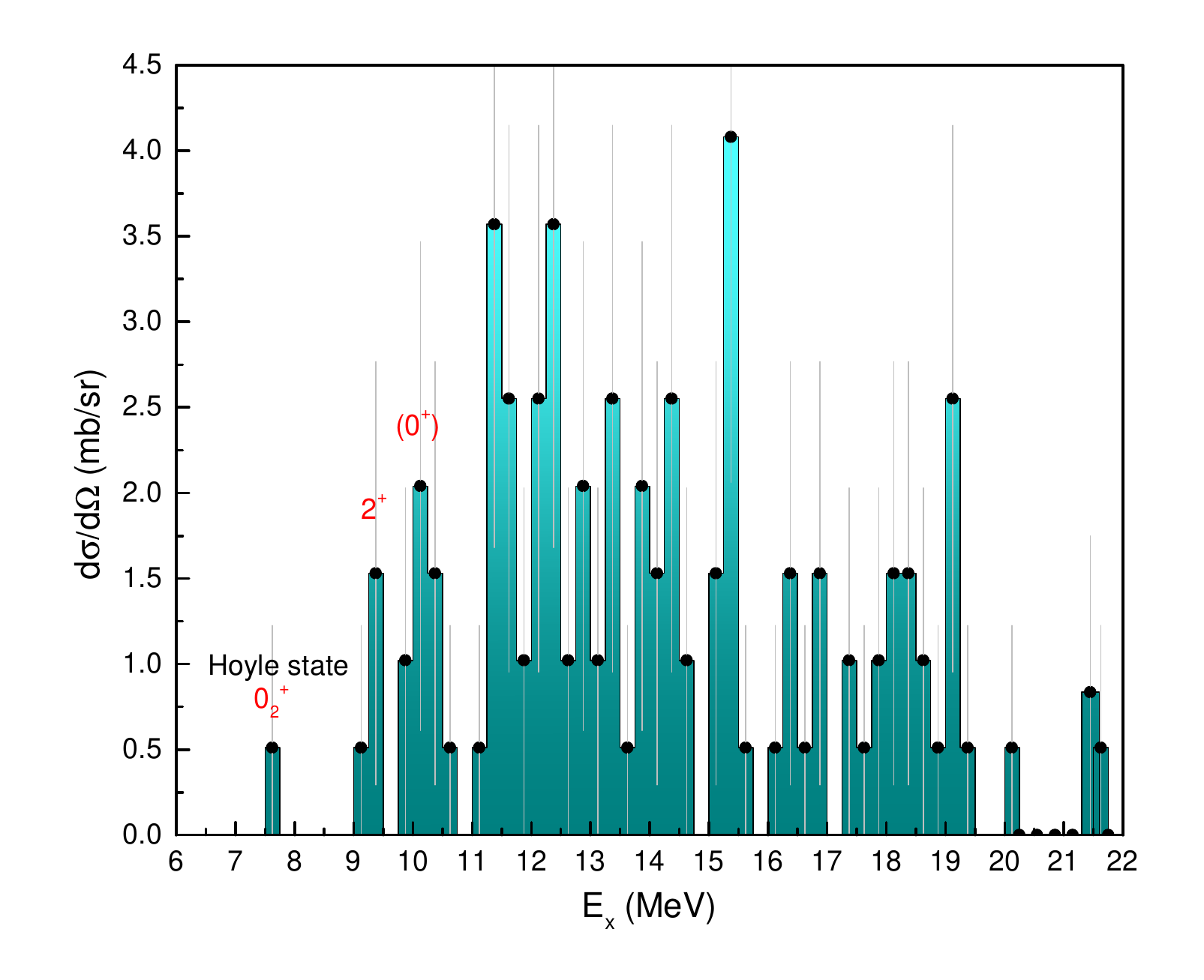}% Here is how to import EPS art
\caption{\label{fig:pdfart} The states of the excitation energy spectrum of $^{12}$C$^\ast$ reconstructed from the invariant mass of three coincident $\alpha$s. }
\end{figure} 

\subsection{Triple $\alpha$s correlation in $^{12}$C}
The angle coverage ($\textless$ 9$^{\circ}$) of the present zero-degree spectrometer indicates that the detected three $\alpha$ particles are emitted in the forewards angle region from the projectile nucleus decay, These three $\alpha$s were kinematicly in forwards direction in lab system from the one $\alpha$ exchange reaction of $^{12}$C($^{11}$C,$^{12}$C$^{\ast}$)$^{11}$C. If these three $\alpha$s are from the excited target nucleus $^{12}$C decay, they are in an almost isotropy emission in the 4$\pi$ space in the lab system, it is difficult to measure them completely by the forewards small angle coveraged detection system. Therefore, the present measurements only can supply for the information of the sequential fusion process of $^{8}$Be$^{\ast}$+$\alpha$ reaction corresponding to the nuclearsynthesis from 2$\alpha$ process to 3$\alpha$ process in astrophysical scenarios. 

It was conjectured that another resonance is at 9-10 MeV with spin-parity of 2$^{+}$ in 1956 [10, 11]. A resonance at 10.1 MeV with a huge broad width of 3 MeV was found soon, while its spin-parity was undetermined by 0$^{+}$ or 2$^{+}$ [11, 20]. In the past several decades little clarification has been brought to this problem, the 2$^{+}$ resonance state at 9.1 MeV with width 0.56 MeV is always adopted in the current compilation of astrophysical reaction rates [11]. The recent measurement observed two resonances at 11.23 MeV with width 2.5 MeV and 13.9 MeV with width 0.7 MeV, but didn't confirm the presence of the resonance at 9.1 MeV, in their work the deduced 3$\alpha$ reaction rate for the temperature from 10$^{7}$ K to 10$^{10}$ K significantly deviates from the standard rates [11].

The excitation energy spectrum of $^{12}$C, as shown in Fig. 7, was generated from the invariant mass of the reconstruction with three coincident $\alpha$ particles. The Hoyle state resonances at 7.65 MeV is the first peak recorded in the excited spectrum. Besides, 9.1 MeV resonance is confirmed at present,  10.1 MeV, 11.5 MeV resonance states and the continuum in the higher excitation energy region can be identified confidently, which are consistent to the generally adopted compilations for the energy levels of $^{12}$C. This excited spectrum reconstructed from the three conincident $\alpha$s indicates that there was some probabilities of the cosmological $^{12}$C productions from the high excited state above Hoyle resonance with an origin of the $^{8}$Be fusion one $\alpha$ nuclear reaction.

The spatial configuration of the excited states produced from 3$\alpha$ reaction process possibly impact the subsequent $\alpha$ process to form the heavier elements. The coupling strength among three $\alpha$s determine the shape of the 3$\alpha$ clusters molecular state. The reflection quantity to probe such intrinsic aspect is essentially the relative momentums $\overrightarrow{P}_{\alpha_{i}\_\alpha_{j}}$=$\overrightarrow{P}_{\alpha_{i}}$-$\overrightarrow{P}_{\alpha_{j}}$, where $\overrightarrow{P}_{\alpha_{i(j)}}$ is the momentum deduced from the measured total energy of $\alpha_{i(j)}$ combined with its emission direction.

\begin{figure}
\includegraphics[width=9.2cm]{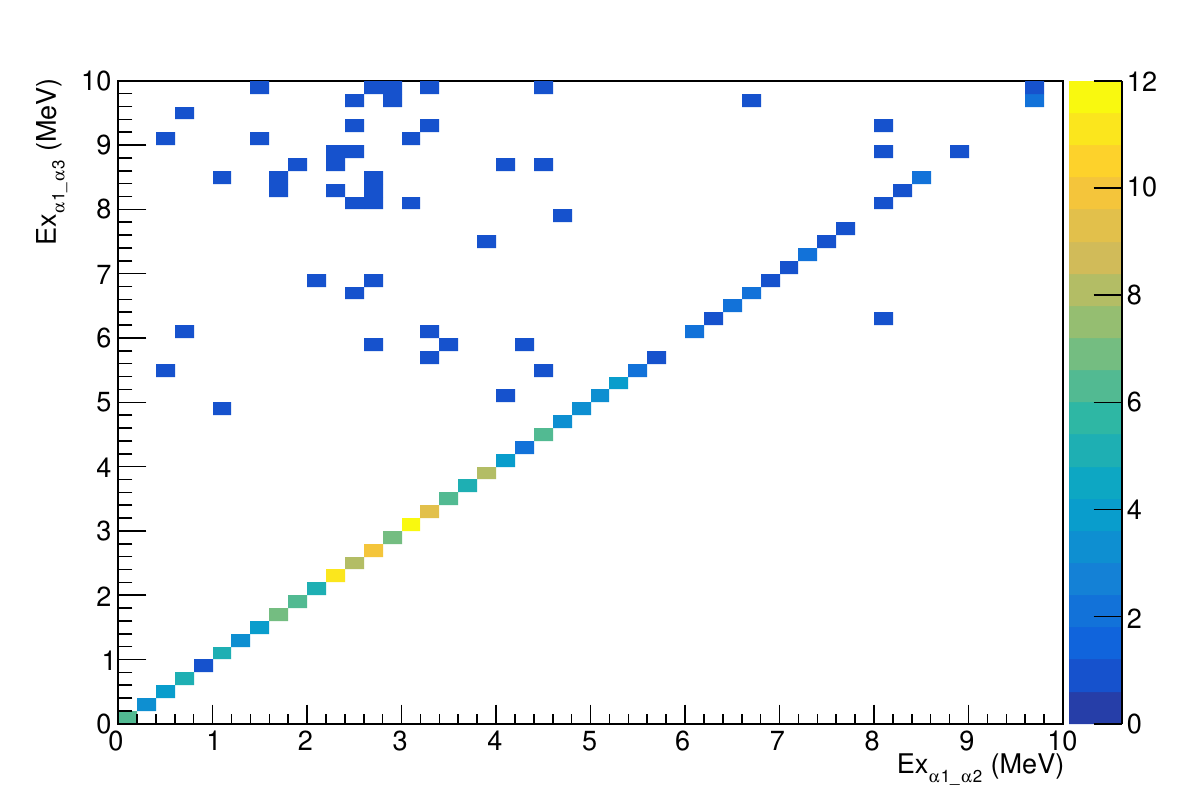}% Here is how to import EPS art
\caption{\label{fig:pdfart}The Dalitz plot of 3$\alpha$s from the decay of the excited states in$^{12}$C, where the invariant masses of $\alpha1$ and $\alpha$2, $\alpha1$ and $\alpha$3 correspond to the excitation energy $Ex$ of $\alpha1$ and $\alpha$2, $\alpha1$ and $\alpha$3 expressed by Equ. (4) and (5).}
\end{figure}

The remarkable property of two identical correlated particles is its relative momentum, around 20 MeV for ($|\overrightarrow{P}_{\alpha_{i}}|$-$|\overrightarrow{P}_{\alpha_{j}}|$)/2 for two correlated protons in the nuclei, which was confirmed by the measured correlation of two protons in the $^{40}$Ca+$^{40}$Ca and $^{48}$Ca+$^{48}$Ca reactions [21], besides, variable calculations based on QMD model [22] have been carried out for constraining the relative momentum of two protons in nuclei.

\subsection{Dalitz plot analysis for 3$\alpha$s decay in $^{12}$C$^{\ast}$}
The Dalitz plot represents a graphical representation of the three alpha correlation in the decay of the excited states of $^{12}$C. It consists of a two-dimensional plot, where the horizontal and vertical axes represent the invariant mass squared of two alpha particles, as shown in Fig. 8. Each point on the plot corresponds to a specific event in the decay process, with the density of points providing information about the probability of observing that particular configuration. By analyzing the distribution of events on the Dalitz plot, ones can extract valuable information about the resonances and decay mechanisms involved in the decay of the excited states of $^{12}$C.

One of the key features of the Dalitz plot is its ability to reveal symmetries and patterns in the decay process. Certain regions around 3 MeV of the plot, known as bands, are associated with specific decay mechanisms $^{12}$C$^{\ast}$ $\rightarrow$ $^{8}$Be$^{\ast}$+$\alpha$ $\rightarrow$ $\alpha$+$\alpha$+$\alpha$ or intermediate resonances of $^{8}$Be. The shape and position of these bands can provide insights into the underlying symmetries of the system, as well as the relative strengths of different decay channels. By comparing the observed Dalitz plot with theoretical predictions and models, ones can test the validity of these models and gain a deeper understanding of the dynamics of the excited states of $^{12}$C.

\section{Discussion}

The formation of such cluster configurations can be understood in terms of a two-body loosely bound system [23]. Assume a square well potential between them, the density distribution of $\alpha$ is $\rho (r)=|\Psi(r)|^{2}\propto \frac{e^{-2\kappa r}}{r^{2}}$, where $\Psi(r)$ is the $\alpha$ wave function. Its slope of the distribution tail is determined by the quantity $\kappa$, which is positively correlated to $\alpha$ separation energy of $E_{s}$ with $(\hbar\kappa)^{2}=2\mu E_{s}$, where $\mu$ is the effective mass of $\alpha$ particle. Binding energy $E_{s}$ of $\alpha$ in excited states of $^{8}$Be$^{\ast}$ is much less than the nearly-constant nucleon separation energy (6-8 MeV) for the ground state of $^{8}$Be, which correspond to the small value of parameter $\kappa$. The momentum distribution of $\alpha$ obtained from the Fourier transform for wave function $\Psi(r)$ can be expressed as $f(p)=C/(p_{i}^{2}+\kappa^{2})$ in the momentum space. The small parameter $\kappa$ correspones to the narrow width of $f(p)$. The distribution in coordinate space is therefore broad with a long tail, if that in momentum space is narrow, according to Heisenberg's uncertainty principle. 

\begin{figure}
\includegraphics[width=9.2cm]{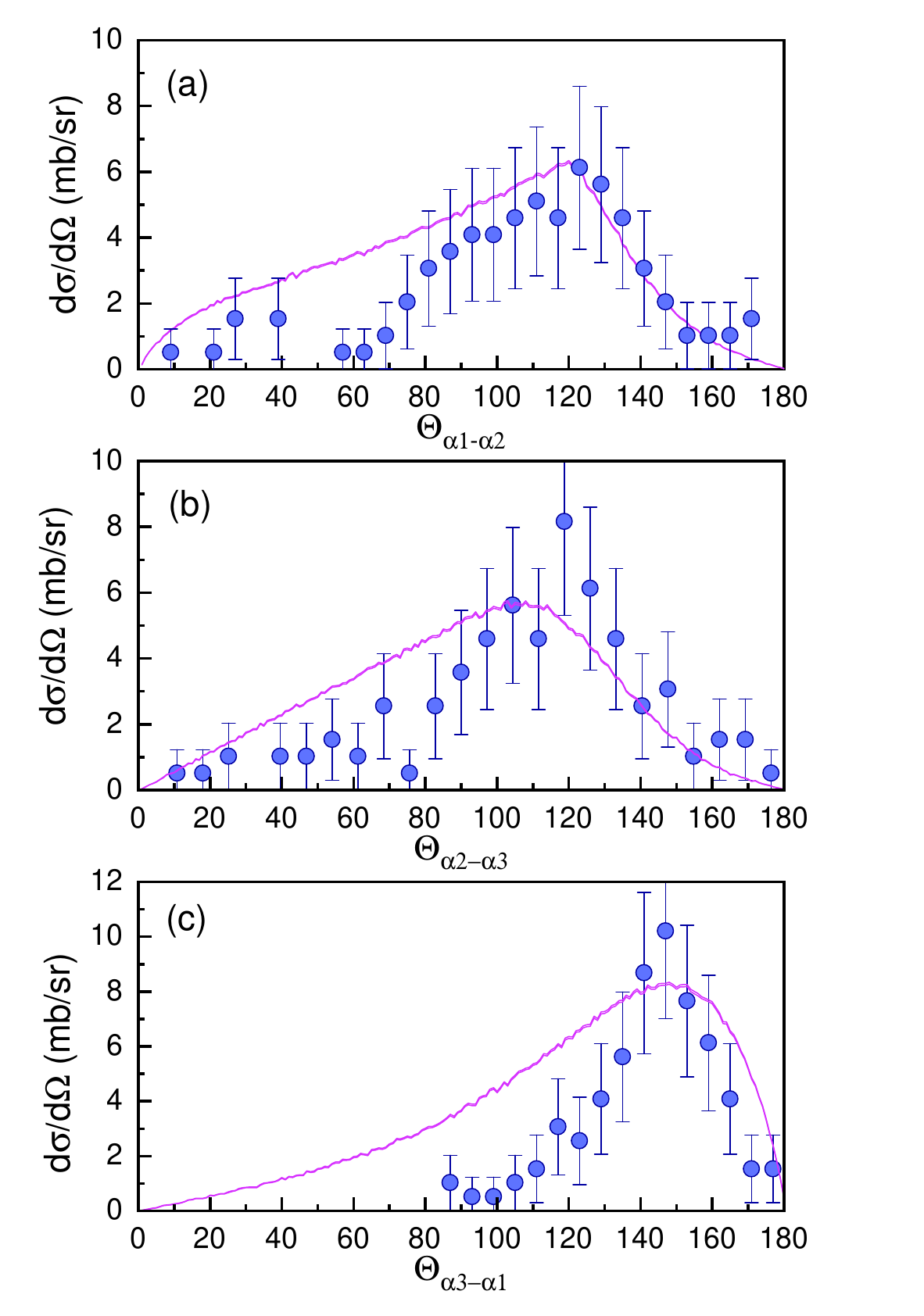}% Here is how to import EPS art
\caption{\label{fig:pdfart}The relative open emission angle $\Theta_{\alpha_{i}\_\alpha_{j}}$ distribution for $\alpha_{1}$$\_$$\alpha_{2}$, $\alpha_{2}$$\_$$\alpha_{3}$ and $\alpha_{3}$$\_$$\alpha_{1}$ interactions. The curves indicate the Monte-Carlo simulations. }
\end{figure}

The Hoyle state is characterized by a strong $\alpha$-$\alpha$ correlation, which arises from the close proximity of the three alpha particles. The $\alpha$-$\alpha$ correlation in the Hoyle state is responsible for its narrow width, which is about 0.4 MeV [18]. The narrow width of the Hoyle state is important for its astrophysical significance, as it ensures that it is efficiently populated in stars. The $\alpha$-$\alpha$ correlation in the Hoyle state can be described in terms of the $\alpha$-$\alpha$ potential, which is the effective potential between two $\alpha$ particles due to the exchange of pions and other mesons. The $\alpha$-$\alpha$ potential has a deep attractive well at short distances, which binds the two alpha particles together. The $\alpha$-$\alpha$ potential also has a repulsive core at longer distances, which prevents the alpha particles from getting too close to each other.

The 2$^{+}$ excited state of $^{8}$Be is another important nuclear state that exhibits a strong $\alpha$-$\alpha$ correlation. The 2$^{+}$ state lies about 3.03 MeV above the ground state of $^{8}$Be and has a spin and parity of $J^{\pi}$ = 2$^{+}$. The 2$^{+}$ state is a resonance state that decays mainly to two $\alpha$ particles. The $\alpha$-$\alpha$ correlation in the 2$^{+}$ state can also be described in terms of the $\alpha$-$\alpha$ potential, which is similar to the potential for the Hoyle state. However, the $\alpha$-$\alpha$ potential for the 2$^{+}$ state is less deep and less attractive than that for the Hoyle state. This is because the 2$^{+}$ state is a less bound state than the Hoyle state and has a larger radius.

Comparison of the $\alpha$-$\alpha$ correlation in the Hoyle state and the 2$^{+}$ state: The $\alpha$-$\alpha$ correlation in the Hoyle state and the 2+ state exhibits many similarities and differences. Both states are characterized by a strong $\alpha$-$\alpha$ correlation, which arises from the close proximity of the $\alpha$ particles. There are some important differences between the $\alpha$-$\alpha$ correlation in the Hoyle state and the 2$^{+}$ state. Firstly, the $\alpha$-$\alpha$ potential for the Hoyle state is deeper and more attractive than that for the 2$^{+}$ state. This is because the Hoyle state is a more bound state than the 2$^{+}$ state and has a smaller radius. Secondly, the Hoyle state is a superposition of three $\alpha$ particles, while the 2$^{+}$ state is a superposition of two $\alpha$ particles. This difference in the number of alpha particles leads to differences in their decay modes and branching ratios. The $\alpha$-$\alpha$ correlation in the Hoyle state and the 2$^{+}$ state also has important astrophysical implications. The Hoyle state is believed to be responsible for the synthesis of carbon and other elements in stars, while the 2$^{+}$ state plays an important role in the triple-$\alpha$ reaction, which is responsible for the production of carbon in stars. The $\alpha$-$\alpha$ correlation in both states affects the rates of these reactions and their branching ratios.

\section{Summary}
Exploring the origin of $^{12}$C, owning to the triple-$\alpha$ process in the stellar helium burning stage via the initial fusion of two $\alpha$ particles ($^{8}$Be) followed by the third one forming the Hoyle state, is extremely fascinating. Taking this resonant into account of the two-step process may successfully explain the observed abundance of carbon in universe. While the width of the first excited $2^{+}$ resonance state of $^{8}$Be was convincingly measured in the present work wider than the compilation quantity, namely the lifetime of the excited $^{8}$Be$^{\ast}$ as the intermediate state from 2$\alpha$ process to 3$\alpha$ process is shorter than the generally expected value, which may lead to the lower aboundance of $^{12}$C formed by the Hoyle state from the $^{8}$Be$^{\ast}$+$\alpha$ fusion process. It is predicted that the complementary $^{12}$C aboundance except from the carbon production from Hoyle state probably contribute from the analogous Hoyle states at higher excited states in $^{12}$C.

The spatial 3$\alpha$ configuration of this analogous Hyole state was confirmed being an isosceles triangle, which indicates the unequal interaction strength and asymmetry couplings among three $\alpha$ constituents at the high excited states in $^{12}$C.

\section{Acknowledgment}

We would like to acknowledge the staff of HIRFL for the stable operation of the cyclotron. The author T. F. Wang appreciates for the financial supports from China Scholarship Council. This work has also been supported by the National Natural Science Foundation of China (No. 10175091 and No. 11305007).

% The \nocite command causes all entries in a bibliography to be printed out
% whether or not they are actually referenced in the text. This is appropriate
% for the sample file to show the different styles of references, but authors
% most likely will not want to use it.

\end{document}